\documentclass[12pt,preprint2]{aastex}

\def\deg{\ifmmode^\circ\else$^\circ$\fi}

\shorttitle{WTTS Disks are Rare}
\shortauthors{Padgett et al.}

\begin{document}
\title{Weak-line T Tauri Star Disks I. Initial {\it Spitzer} Results from the Cores to Disks Legacy Project}   

\author{Deborah L. Padgett\altaffilmark{1}, Lucas Cieza\altaffilmark{2},
Karl R. Stapelfeldt\altaffilmark{3}, Neal J. Evans, II\altaffilmark{2},
David Koerner\altaffilmark{4}, Anneila Sargent\altaffilmark{5},
Misato Fukagawa\altaffilmark{1}, Ewine F. van Dishoek\altaffilmark{6},
Jean-Charles Augereau\altaffilmark{6},
Lori Allen\altaffilmark{7}, Geoff Blake\altaffilmark{5}, Tim Brooke\altaffilmark{5}, 
Nicholas Chapman\altaffilmark{8},
Paul Harvey\altaffilmark{2}, Alicia Porras\altaffilmark{7},
Shih-Ping Lai\altaffilmark{8}, Lee Mundy\altaffilmark{8}, 
Philip C. Myers\altaffilmark{7}, William Spiesman\altaffilmark{2}, 
Zahed Wahhaj\altaffilmark{4}}
\altaffiltext{1}{{\it Spitzer} Science Center, MC220-6, California Institute of Technology,
    Pasadena, CA 91125 dlp@ipac.caltech.edu}
\altaffiltext{2}{Astronomy Departement, University of Texas, 1 University
Station C1400, Austin, TX 78712}
\altaffiltext{3}{Jet Propulsion Laboratory, MS 183-900, 4800 Oak Grove Drive,
Pasadena, CA 91109}
\altaffiltext{4}{Northern Arizona University, Department of Physics and Astronomy,
Box 6010, Flagstaff, AZ 86011}
\altaffiltext{5}{California Institute of Technology, Pasadena, CA 91125}
\altaffiltext{7}{Smithsonian Astrophysical Observatory, 60 Garden Street,
MS42, Cambridge, MA 02138}
\altaffiltext{8}{Astronomy Departement, University of Maryland, College Park, MD
20742}
\altaffiltext{6}{Leiden Observatory, Postbus 9513, 2300 R.A. Leiden, Netherlands}
\begin{abstract} 

Using the {\it Spitzer Space Telescope}, we have observed 90 weak-line and classical
T Tauri stars in the vicinity of the Ophiuchus, Lupus, Chamaeleon, and
Taurus star-forming regions as part of the
Cores to Disks (c2d) {\it Spitzer} Legacy project. In addition to the
{\it Spitzer} data, we have obtained contemporaneous optical photometry
to assist in constructing spectral energy distributions. These objects were specifically
chosen as solar-type young stars with low levels of H$\alpha$ emission,
strong X-ray emission, and lithium absorption i.e. weak-line T Tauri stars, most of 
which were undetected
in the mid-to-far IR by the IRAS survey. Weak-line T Tauri stars are potentially extremely
important objects in determining the timescale over which disk
evolution may take place. Our objective is to determine whether these
young stars are diskless or have remnant disks which are below
the detection threshold of previous infrared missions.
We find that only 5/83 weak-line T Tauri stars have detectable  
excess emission between 3.6 and 70 \micron\ which would indicate the presence of 
dust from the inner few tenths of an AU out to 
the planet-forming regions a few tens of AU 
from the star. Of these sources, two have small excesses
at 24 microns consistent with optically thin disks; the
others have optically thick disks already detected by
previous IR surveys.  All of the seven classical T Tauri stars show
excess emission at 24 and 70 \micron\, although their
properties vary at the shorter wavelengths. Our initial results 
show that disks are rare among young 
stars selected for their weak H$\alpha$ emission. 

\end{abstract}

\keywords{stars: pre-main sequence ---  (stars:) planetary systems: protoplanetary disks ---  infrared: stars}

\section{Introduction}

T Tauri stars (Joy 1945) have long been known as solar-mass
pre-main sequence stars. Sprinkled throughout molecular cloud complexes,
or gathered in dense clusters, they have traditionally been
identified by their unusual optical spectra which include
bright emission lines of H$\alpha$ and other permitted and
forbidden atomic transitions. However, with the advent of
X-ray molecular cloud surveys, another class of solar-type
young star has been identified from their high X-ray luminosity.
Unlike the spectroscopically identified ``Classical" T Tauri stars (CTTS),
these objects lack H$\alpha$ emission in excess of 10 \AA\ in
equivalent width (EW)
and are thus known as ``weak-line" T Tauri stars (WTTS; 
Herbig \& Bell 1988).
Confirmation that X-ray identified WTTS are young stars
requires high resolution optical spectra
that show a solar-type spectrum with a Li I 6707 \AA\ 
absorption line equivalent width stronger than that of a Pleiades
star of the same spectral type (Covino et al. 1997). The presence
of a strong lithium absorption indicates stellar youth 
among stars with deeply convecting photospheres since
lithium is easily destroyed at high temperatures.
In practice, using the Li I EW to select young stars 
is a difficult task, since the strength of the lithium
line may depend on the rotational history of the star and spot
coverage (Mendes et al. 1999). In addition, spectral classes earlier than
G5 may retain strong lithium lines
throughout their main sequence lifetimes (Spite et al. 1996). Thus, current
lists of WTTS are probably contaminated to some degree by
young main sequence stars which are closer than the cloud,
but too faint for accurate parallax measurements (Alcala et al. 1998). 

\par
The current paradigm of planet formation is still based
on the venerable IRAS (Strom et al. 1989) and millimeter continuum
surveys (Beckwith et al. 1990) of the late 1980's that
found mid-to-far-infared and/or millimeter excess emission around 
about half of solar-type stars in nearby clouds. This picture
of disk evolution starts with Class 0-I sources which are extremely 
faint shortward of 25 microns, proceeds through Class I with a rising
IR Spectral Energy Distribution (SED), continues to 
Class II which has a flat or falling SED with strong IR
excesses above the photosphere, and ends at Class III
where the excess emission indicates a tenuous, optically
thin disk (Adams, Lada, \& Shu 1987; Andr\'e \& Montmerle 1994).
In practice, IRAS and ISO were not sensitive
enough to detect optically thin disks in the mid-IR at the distance
of nearby star formation regions; however, the {\it Spitzer
Space Telescope} does have this capability.
Almost as soon as the first WTTS were identified, they 
were presumed to be the Class III ``missing link" of
the planet formation paradigm since ``post" T Tauri
stars had proved elusive. Initial ground-based NIR and MIR
studies showed that inner disks were very rare among
the WTTS population (Skrutskie et al. 1990). More 
recent JHKL studies of several young clusters 
indicated that inner disks disappear in 5 - 6 Myr (Haisch, 
Lada, \& Lada 2001).  On the other hand, a JHKL study of
the $\approx$ 10 Myr $\eta$ Cha cluster found a rather
high disk fraction among its few known members (Lyo et al.
2003), although a smaller percentage of disks
was found by Haisch et al. (2005).  Both IRAS and 
millimeter surveys indicate that optically
thick outer disks are rare among WTTS (Strom et al. 1989,
Osterloh \& Beckwith 1995). However, a sensitive study 
of tenuous material in the terrestrial planet 
forming zone of WTTS has awaited the power of the
{\it Spitzer Space Telescope} ({\it Spitzer}).
{\it Spitzer} has over 1000 times the sensitivity
at 24 \micron\ of the IRAS satellite, with similar
improvements in its other imaging bands (Werner et al. 2004).
Only {\it Spitzer} can distinguish between transitional disks
(optically thick with inner holes), optically thin
remnant disks with low disk luminosity L$_d$ (L$_d$/L$_*$ $<<$ 
1, like the debris disks found
around older stars), and completely diskless ``naked"
T Tauri stars.

   Study of circumstellar material around WTTS promises
to answer some significant questions about planet 
formation around low mass stars. The first is whether
some young stars are born ``diskless"; that is, some
stars at the age of 1 - 3 Myr do not 
retain remnant circumstellar material and did not
possess substantial disks long enough
for plausible planet formation. Typical ``core accretion"
models of giant planet formation require a minimum
of 5 million years to make a Jupiter, most of
which is spent in the coagulation of dust and
accretion of planetesimals (Wiedenschilling 2000), although
accretion times continue to drop in the most recent
models (Goldreich et al. 2004a,b) and the alternative
``disk instability" model can produce giant planets on
a very short timescale (Boss 2001). 
A low detection frequency of any infrared excess among
WTTS would provide evidence for diskless young stars.
In addition, this study will address the dispersion in disk properties
among coeval stars in the same environment. These differences
could possibly be related to the early formation of planets
in the stars where the disks have inner holes or have
disappeared entirely. Such a hypothesis is
potentially testable with future high resolution astrometric studies 
of young stars with and without disks to see if the presence
of giant planets is related to the state of disk evolution.

Several {\it Spitzer} surveys to assess the detection frequency
of excess in young stars are either in progress or have been
published recently. The entire range of $Spitzer$ instrumentation
are utilized by these studies, including the 3.6 - 8.0 \micron\
camera (IRAC), the 24, 70, and 160 \micron\ photometer (MIPS),
and the  5 - 40 \micron\ spectrograph (IRS).
There are several investigations of excess 
among very young stars in individual star-forming regions (Reach et al. 2004,
Muzerolle et al. 2004, Padgett et al. 2004,
Young et al. 2005, Hartmann et al. 2005a, etc.). These studies 
investigate known young star populations including WTTS, 
CTTS, and Class I sources. Among older pre-main sequence associations,
Chen et al. (2005) report on a MIPS 24 \& 70 \micron\
survey of 40 F- and G- type stars in the Sco-Cen OB association
with an age of 5 - 20 Myr (Chen et al. 2005).
The Megeath et al. (2005) IRAC study of Eta Cha
covers the stellar population of this 8 - 10 Myr association.
A similar investigation with MIPS of the TW Hydrae group
is detailed in Low et al. (2005). Young main sequence clusters
such as M47 (Gorlova et al. 2004) and the Pleiades (Stauffer
et al. 2005) have been studied using MIPS.  For older Population 
I stars, Beichman et al. (2005a) and
Bryden et al. (2006) present results from a very sensitive 24 \micron,
70 \micron (MIPS), and {\it Spitzer} spectrograph (IRS)
survey of generally old, but very bright, solar neighborhood stars.
Due to the brightness of their targets and selection for low
IR backgrounds, this program has much higher sensitivity at
70 \micron\ than the other studies, enabling detection of very
tenuous cold disks invisible to other surveys.  Two surveys which span a 
wide range of stellar ages are Rieke et al. (2005) and
the ``Formation and Evolution of Planetary Systems'' 
$Spitzer$ Legacy project or FEPS (Meyer et al. 2004). For stars
more massive than the sun, Rieke et al. (2005) surveys A-type stars of
5 - 850 Myr at 24 $\micron$ with MIPS.
FEPS is examining stars with ages ranging from 3 Myr - 3 Gyr, divided
into bins for stars of roughly equivalent ages (Meyer et al.
2004). The large age span for FEPS means that their sample has a
large number of faint young stars at distances over 100 pc and a
smaller number of bright nearby old stars for which 70 \micron\
sensitivities are good. Although their youngest age group includes solar-type stars of 
similar age to the current
WTTS sample, the FEPS young stars are located more than 6$\deg$
from nearby clouds (Silverstone et al. 2006).  In contrast to these 
surveys,  our 0.5 - 70 \micron\ survey focuses only on WTTS in and adjacent to several
star-forming clouds within 200 pc. 

The current work is the first of three papers that will
discuss disks around WTTS in the c2d {\it Spitzer}
data. Paper I discusses initial results for 90 stars specifically
targeted for photometric mode observations with IRAC and MIPS. 
Most of these sources fall
outside of high extinction regions but within 6 degrees of cloud
centers.  Paper II (Cieza et al.
2006) will present SEDs for WTTS that fall within the
large c2d cloud maps of Perseus, Ophiuchus, Lupus, Chamaeleon,
and Serpens. Paper III (Wahhaj et al. 2006) will present
the remainder of the individually targeted WTTS, 
provide upper limits for the 70 micron non-detections,
and interpret the entire sample in terms of debris disk models.

\section{Observations}
\par
\subsection{Target Selection}
The c2d WTTS survey source selection focuses on targets associated with nearby
molecular clouds being mapped by large {\it Spitzer} mapping surveys.
Only for nearby targets (d $<$ 200 pc) can {\it Spitzer} achieve 
the needed sensitivity to optically thin dust in the planet formation region 
around a ZAMS low-mass star, over a statistically meaningful sample, and in the 
50 hours of observing time allocated for this study.  Molecular cloud
association is crucial for establishing a star's youth: over 10 Myrs,
the measured proper motion dispersion for young stars at d$\sim$ 140 pc
can carry them 6$^{\circ}$ of projected separation away from their formation
region (Hartmann et al. 1991).  We therefore required that our WTTS targets 
lie within this radius from their associated cloud.  About 1/3 of the
sample lies within the clouds, 1/3 are adjacent to the clouds,
and 1/3 are 3 - 6 degrees from the clouds. Selecting WTTS associated 
with clouds being mapped by {\it Spitzer} also allows an eventual comparison 
between the WTTS disk properties and those of the younger cloud population;
both presumably formed under similar conditions.  In the end, rich but
distant regions of low-mass star formation (Orion, Serpens, Perseus) were 
rejected for study in favor of those in the distance range 120-160 pc: Taurus, Lupus, 
Ophuichus, and Chamaeleon.  The latter three are being mapped by the c2d 
project (Evans et al. 2003), and a separate large {\it Spitzer} survey of 
Taurus is also being carried out (G\"udel, Padgett, \& Dougados 2006). 

The ROSAT mission of 1990-1998 identified a sizable population
of X-ray active stars in and around the four nearby clouds targeted for 
this WTTS study (Wichmann et al. 1996, 1997). X-ray activity is a signature 
of stellar youth, and is correlated with the higher rotational velocities found 
as young stars contract down toward the main sequence.  Follow-up observations 
on the ROSAT samples in these clouds (Wichmann et al. 2000, Taurus; Covino et al. 
1997, Chamaeleon; Wichmann et al. 1999, Lupus; and Mart\'in et al. 1998, 
Ophiuchus) verified
the presence of strong Li I 6707 \AA\ absorption in the ROSAT samples.  
The {\it Spitzer} WTTS targets were selected from these studies, requiring that 
the object show Li absorption stronger than that seen in 10$^8$ Myr Pleiades stars 
and have a spectral types G5 or later where 
the Li-age relationship is best understood.  Brighter objects were
favored, as more of them could be observed at the program design 
sensitivity (see below).  WTTS discovered by earlier surveys in H$\alpha$ 
emission, Ca II emission (Herbig \& Bell 1988, and by the $Einstein$ satellite (e.g., 
Walter et al. 1986) 
were also included in the sample. H$\alpha$ equivalent widths for the
WTTS sample range from 9\AA\ (in emission) to $\sim$ 2\AA\ in
absorption. Only about 5\% of the sample has H$\alpha$ in absorption.
In addition, seven 
stars with CTTS H$\alpha$ characteristics and confused IRAS-based
SEDs were also included. These are treated
separately in our statistics, forming a useful sample
of stars with spectroscopic disk indicators for comparison
with our WTTS.
A total of 188 objects, spanning
spectral types G5-M5, were selected.  Results for the first 90
stars of this sample are reported here; the remainder will be discussed
in Papers II and III.

\subsection{{\it Spitzer} Photometry}

The c2d WTTS observations were carried out with both {\it Spitzer's} 
Infrared Array Camera (IRAC; Fazio et al. 2004) and Multiband Imaging 
Photometer (MIPS; Rieke et al. 2004).  Target sensitivities for the
observations were defined relative to photospheric flux densities;
these were estimated by extrapolating optical and near-IR photometric 
measurements into the mid-IR using a blackbody appropriate to each star's 
spectral type.  The IRAC observations consisted of two 12-sec exposures 
with an offset dither, plus two short 0.6 sec exposures (``HDR mode''), 
taken in both camera fields of view.  These achieved sensitivities sufficient 
to detect the photospheres of all the targets at S/N $\ge$ 50 in IRAC's 
broad bands centered at 3.6, 4.5, 5.8, and 8.0 $\mu$m.  The MIPS 24 $\mu$m 
photometry observations were designed to detect each star's photosphere 
at S/N$=$ 20.  Total exposure times ranging from 42-420 sec were used, 
depending on the brightness of each individual target.  At 70 $\mu$m,
the typical stellar flux density of 6 mJy is too faint to readily
detect.  Most targets were therefore observed at a fixed total exposure
time of 360 sec.  This was sufficient to detect a disk like that of
beta Pictoris (L$_d$/L$_*$= 10$^{-3}$; excess 10$\times$ the photosphere) 
in the brighter half of the sample, and disks 3-10 times more luminous 
than this in the other half. Thus, only half of our sample has
observations with enough 70 \micron\ sensitivity to detect a cold $\beta$
Pic-like disk. 

A list of the targets and observation dates is given in Table 1.
Some observations were made in a ``cluster'' mode combining the
datasets from objects located within 1$^{\circ}$ of each other on the
sky.  In such cases, the data in the {\it Spitzer} Science Center 
Archive can be found by referencing the name of the first target
in the cluster.  The c2d WTTS data are indexed under {\it Spitzer} 
observing program 173, moniker ``COREPLANETS\_STARS''. Enhanced 
versions of the data products provided by the c2d team can be
found at http://data.spitzer.caltech.edu/popular/c2d/.

Following the observations, the data were processed through the SSC
data reduction pipeline version S11.0. The c2d team then applied
a number of algorithms designed to mitigate remaining instrumental
signatures (Young et al. 2005). The WTTS were
located on the images using 2MASS positions, which in some
cases are offset by 20 - 40$\arcsec$ from the ROSAT positions
reported in the literature. Pointing for IRAC is offset from 2MASS
by a fraction of an arcsec, while MIPS 24 $\micron$ pointing is
reliable to within 2$\arcsec$ (Fazio et al. 2004; Rieke et al. 2004).
PSF-fitting photometry was performed on the mosaicked images using
the ``cd2phot" code adapted by the c2d team (Young et al. 2005;
Harvey et al. 2006). Photometric measurement uncertainties of about 2\%
are typical for moderately bright IRAC and MIPS-24 sources. The
absolute calibration errors are estimated at approximately 5\% for IRAC and 
7\% for MIPS-24 (SSC Data Handbook). Unlike the other bands,
photometry for MIPS-70 detections have been measured using aperture
photometry on the filtered SSC post-BCD mosaics.  For these preliminary 
photometric measurements, uncertainties are estimated at 30\%. 
A forthcoming paper will contain improved
70 \micron\ photometry and rigorous upper limit estimates for the 
70 \micron\ fluxes of these sources.  Bright sources in any of the bands 
may have a higher uncertainty due to PSF-fitting problems and latent image effects.

\subsection{Optical Photometry}
In order to construct SEDs using contemperaneous optical, NIR,
and mid-IR data, 
VRI optical observations of the Taurus objects were made with the 0.8 m
telescope at McDonald Observatory on 5 different photometric
nights during 2 observing runs in December 2003 and January 2004.
The southern targets were observed with the 0.9 m  telescope at Cerro
Tololo Inter-American Observatory during 4 photometric nights in May
2004. In addition to the program stars, on each night, 3 fields of
Landolt standards ($\sim$ 5 standards per field) were  observed
at different airmasses. We tailored our exposure times to obtain
excellent S/N in every observation. The standard IRAF packages CCDRED
and DAOPHOT were used to reduce the data and perform  aperture
photometry. The rms scatter of the photometric solutions applied to
the programs stars was 0.01-0.02 mag in all three filters.
We adopt a conservative photometric error of 0.02-0.03 mag.

\section{Results}
\subsection{Frequency of Excess Emission}
\par
Due to uncertainties
in extinction, spectral types, and absolute photometry, relative
photometric colors are used for detecting excess
emission. This technique has been widely utilized by
{\it Spitzer} researchers to reveal IR excesses in young
stars (Gorlova et al. 2004; Rieke et al. 2005;
Beichman et al. 2005a).  We have chosen to identify
excesses at 8 and 24 microns using the K$_s$ - [24]
and K$_s$ - [8] color. These colors are approximately
zero for Rayleigh-Jeans photospheres in spectral types
earlier than late M (Gorlova et al. 2004). Figure 1 shows
a plot of 2MASS K$_s$ magnitude versus K$_s$ - [24] color
for our 90 star sample.
The sample is dominated by the locus of points with
K$_s$ - [24] approximately zero. This group of sources
spans brightnesses from K$_s$ = 8 to 11, covering our
range of spectral types in nearby clouds. The width of this 
distribution is $\approx$ 0.2, indicating the minimal
photometric errors in our dataset. 
Substantial K band extinction
might also tend to produce red K$_s$ - [24] colors.
However, published A$_v$ for these stars indicate
very low NIR extinctions for most of the sources. In addition,
Figure 2 plots H - K$_s$ versus K$_s$ - [24],
and the reddening vector for A$_v$ = 5 demonstrates that even 
high extinction cannot produce excess sources on our 
color-magnitude diagram since the principal displacement is
perpendicular to the K$_s$ - [24] axis. 
Twelve sources are located from 1 - 3 magnitudes redward of
the photospheric locus. These are the stars for which we detect
excess emission at 24 microns. Seven of these stars
are CTTS; only five are WTTS. Interestingly, the 
only sources obviously detected at 70 microns are
the ones with 24 micron excesses. Thus, unlike surveys
of older stars in the solar neighborhood for which
70 $\micron$ excesses are fairly common (Bryden et al. 2006), 
we have thus far found no young stars with obvious excess only at 
$\lambda$ $\geq$70 $\micron$.
However, we emphasize that unlike our observations,
the Bryden et al. (2006) 70 \micron\ sensitivity limit reaches the
stellar photosphere for all their targets. 
Figure 3 presents a similar plot for K$_s$ - [8.0].
Out of 90 stars, nine stars (3/83 WTTS; 6/7 CTTS) show excess at 
8 \micron. Three stars with 24 micron excess lack 8 $\micron$
excess. Of these, two are WTTS. The third is the low luminosity
CTTS Sz 84. Our minimal group of 7 CTTS has a 100\% excess detection rate
at 24 and 70 microns. 
On the other hand, only 5 out of 83 = 6\% of the stars with H$\alpha$ 
EW $\leq$ 10 \AA\ in emission have 
a detectable excess at 24 or 70 microns. SEDs for WTTS with excess
emission are presented in Figure 4. SEDs for CTTS may
be found in Figure 5.
These SEDs are constructed from our optical VRI photometry, 
2MASS JHK photometry, {\it Spitzer} IRAC (3.6, 4.5, 5.8, 8.0
\micron), and {\it Spitzer} MIPS (24, 70 \micron).
Note that the photospheric models have been normalized to
I band.
\subsection{Fractional Disk Luminosities}
\par
Table 3 contains fractional disk luminosities L$_d$/L$_*$ for
the excess sources. Within our sample of excess objects,
L$_d$/L$_*$ ranges from 0.19 - 0.008. Our CTTS disk values
range from 0.19 - 0.05. On the
other hand, the famous debris disk $\beta$
Pic has a fractional luminosity of only 0.002 (Backman \& Paresce
1993). One unusual CTTS (Sz 84) has an exceptionally small
L$_d$/L$_*$ $<$ 0.001. It is discussed in Section 3.3.6. Three of
the WTTS with disks have luminosities consistent with the
lowest part of the CTTS range.  The remaining two WTTS disks are almost an
order of magnitude less luminous than the other sources. Their
L$_d$/L$_*$ falls squarely between the Class II and debris
disk range, filling an important gap in our knowledge of
disk evolution. 

\subsection{Objects with Excess Emission} 
\subsubsection{RXJ0432.8+1735}
RXJ0432.8+1735 is a ROSAT-detected star near the Taurus
star formation region at a distance of 140 pc. 
This object was first reported by Carkner et al. (1996)
as a strong X-ray source in an 8 ks pointed ROSAT
observation that was subsequently identified as 
an M-type weak H$\alpha$ emission star in the L1551 cloud. 
Wichmann et al. (1996) reported a 
spectral type of M2 and an H$\alpha$ equivalent 
width  of 1.7 \AA\ in emission. Undetected to a
limit of 0.19 mJy at 8.4 GHz in a radio survey 
by Carkner et al. (1997), RXJ0432.8+1735
was also observed using speckle interferometry by Kohler
\& Leinert (1998), who found no companions
down to $\Delta$K $\approx$ 3 at 0.13$\arcsec$.
Marti\'n \& Maguzzu (1999) obtained high resolution
spectroscopy, confirming this star's spectral type
and low H$\alpha$ EW, as well as showing that its
lithium EW places it squarely in the WTTS regime. 
Hanson et al. (2004) determined proper motions of
$\alpha$ = 11.3 mas/yr, $\delta$ -19 mas/yr; a similar value was
found by Ducourant et al. (2005), consistent with
the CTTS GG Tau, which is only 7 arcmin away. 
The proximity of RXJ0432.8+1735 to this 1.5 Myr circumbinary 
disk source also provides evidence of its young age (White
et al. 1999). The pre-main sequence tracks of D'Antona
\& Mazzatelli (1997) yield an age of 1$\pm$0.5 Myr. 
The SED is presented in Figure 4.
All of the {\it Spitzer} photometry points fall very close to the
photosphere with the exception of MIPS 24 $\micron$.
The 24 $\micron$ flux is a factor of 3 above
the expected photospheric value and is detected
at a high S/N level. There was no detection of MIPS 
70 $\micron$ for this source.
\subsubsection{RXJ1603.2-3239}
RXJ1603.2-3239 is a young star in the Lupus star forming
region at a distance of $\sim$150 pc (Sartori et al. 2003). 
This object was first reported as a K7 X-ray bright star 
with H$\alpha$ EW = 1.13 \AA\ in emission (Krautter et al. 1997).
The star has a photometrically determined rotation
period of 2.77 days with an amplitude of 0.11 mag
in V (Wichmann et al. 1998). Shortly thereafter,
Wichmann et al. (1999) performed high resolution
spectroscopy on this star, confirming its lithium
EW in excess of Pleiades values for its spectral
class and showing a significant central absorption in its
H$\alpha$ emission which indicates a non-chromospheric
origin for this line. Sartori et al (2003) report A$_v$ = 0.49 
and derive log L$_*$ = -0.40 based on a distance for nearby 
Hipparchos-detected early-type stars. The D'Antona \& 
Mazzatelli (1997) tracks indicate an age of 1.7$\pm$0.4 Myr.
Our SED (Figure 4) shows all IRAC
photometric points lying near the photosphere.
However, the 24 $\micron$ flux
is a factor of three above the predicted photospheric
value. Like RXJ0432.8+1735, this source appears to
have an SED consistent with a tenuous disk that
has a large inner hole. 
\subsubsection{V836 Tauri}
V836 Tauri is a K7 star in the western part of the
Taurus star forming region near L1544.
A thorough review of its optical photometric and
spectroscopic properties is presented in 
White \& Hillenbrand (2004). Its H$\alpha$ EW in that
work is cited as 25 \AA\, although the HBC (Herbig
\& Bell 1988)
gives the emission strength as 9 \AA\ (Mundt et al. 
1983). An even lower value of 4.4 \AA\ 
with an inverse P Cygni profile was
reported by Wolk \& Walter (1996). 
These measurements classify V836 Tau as a 
borderline WTTS/CTTS.  V836 Tauri and LkCa 15
were the only WTTS detected in the millimeter continuum
and CO survey of Duvert et al. (2000) and is one of
the few WTTS detected in the millimeter continuum
by Osterloh \& Beckwith
(1995). V836 Tauri is sometimes claimed as an ``older"
T Tauri star due to derived ages in excess of
7 Myr (Strom et al. 1989); however, Hartmann (2003) 
casts some doubt on the accuracy of such 
model dependent estimates.
V836 Tauri has a faint nearby companion that 
is likely a background star (Massarotti et al. 2005).
Despite the weak H$\alpha$ emission, IRAS strongly detected
excess emission from this star at all 4 bands (Wolk
\& Walter 1996, Kenyon \& Hartmann 1995).
Our dereddened SED (Figure 4) shows a modest
amount of excess emission in the NIR, almost none
at 3.6 \micron\, and an increasing amount in the
IRAC and MIPS bands. The excess emission in this
source has been fit with an optically thick 
disk, most recently by Andrews \& Williams (2005),
who found a disk mass of 0.01 M$_\odot$. Thus,
although V836 Tau has strongly variable H$\alpha$,
it has excesses typical for a CTTS, although it
shows little excess in the NIR bands.
\subsubsection{Sz 82}
Sz 82, also known as IM Lupi, is a M0 star in the Lupus
star forming region. The H$\alpha$ emission
is known to vary from 7.5 - 21.5 \AA\ confirming 
its status as a borderline WTTS/CTTS. Although Sz 
82 has not been observed by a millimeter interferometer, 
it is detected strongly in the 1.3 mm continuum
(260 mJy; N\"urnberger et al. 1998). The Sz 82 disk 
has been resolved in scattered light by HST (Padgett 
et al. 1999, 2005).  Like V836 Tau, this source was
detected by IRAS (Carballo et al. 1992). Our 
SED (Figure 4) shows a modest NIR excess
that begins to increase at 5.8 microns
and is nearly flat between 8 and 70 microns.
A simple disk model for this object's SED is presented
in Section 4.2.
\subsubsection{Sz 76}
Sz 76, an M1 star in Lupus, is another
borderline WTTS/CTTS with an H$\alpha$
EW = 10.3 \AA\ (Hughes et al. 1994). 
The source was detected in X-rays during a
pointed ROSAT observation of Lupus I
(Krautter et al. 1997). It was not seen
at 1.3 mm down to a detection limit of 45 mJy
which constrains its cold disk mass to $\leq$
5$\times10^{-3}$ M$_{\odot}$ (Nu\"rnberger et al. 1997).
Although no IRAS detections have
been reported for this source due to this mission's
limited sensitivity, we find substantial
excess emission for this star longward of 4.5
microns (Figure 4). A preliminary effort to
model the possible disk of Sz 76 is reported in 
Section 4.2. 
\subsubsection{Classical T Tauri Stars}
Seven CTTS either without IRAS detections or
with confused IRAS SEDs were included
in our sample. These vary in H$\alpha$
properties from the borderline WTTS/CTTS
Sz 77 (EW(H$\alpha$) = 12 \AA; Hughes et al 1994) 
to RXJ1556.1-3655 (EW(H$\alpha$) = 83 \AA\
Wichmann et al. 1999). 
The SEDs of the CTTS are presented in Figure 5 with
the exception of Sz 77, which is in Figure 4. 
A wide variety of SED morphologies are displayed by these sources.
The most extreme levels of excess emission from 
1.2 to 70 microns are shown by Wa Oph/6, a K6 star 
with EW(H$\alpha$) = 35 \AA\ (Walter et al. 1986). 
On the other hand, Sz 84, an M5.5 star with EW(H$\alpha$)
= 44 \AA\, has almost no measureable excess
shortward of 24 microns, implying a sizeable
disk inner hole. The K5 star RXJ1615.3-3255 (EW(H$\alpha$)
= 19 \AA) also has very little excess shortward
of 8 microns. Thus, within our very limited sample,
H$\alpha$ equivalent widths do not appear to correlate with 
the presence of infrared excess, although it
seems that IR excess is rare among stars with 
H$\alpha$ EW $\leq$ 10 \AA.
This finding appears to extend to somewhat
below the WTTS defining level of 10 \AA\ H$\alpha$
emission since the three borderline WTTS/CTTS in
our sample show disk emission that is indistinguishable
from the CTTS in our sample. 
\par
The purported CTTS Sz 84 merits additional consideration
due to its very low fractional disk luminosity of $<$ 0.001.
Because of its relatively high H$\alpha$ equivalent width,
this source was listed as a Lupus T Tauri star
in Schwartz (1977).  However, it was not included in Herbig \& Bell
(1988) despite being discovered well in advance of
this seminal work. Its long pedigree and faint visual
magnitude of 16.2 (Hughes et al. 1994) has apparently caused it to be
overlooked entirely in Li I 6707 \AA\ equivalent width
determinations. Sz 84 also stands out as the star with the
lowest N$_H$ determined from {\it ROSAT} observations
in Krautter et al. (1997). It has a column density of hydrogen
that is considerably lower than the
bulk of stars in Lupus (log N$_H$ $\sim$ 18 versus 21). Although
its estimated age on the D'Antona \& Mazzatelli (1997)
tracks is 1.0$\pm$0.3 Myr, its age is entirely dependent
on placing the object at the distance of the cloud. A
young main sequence star in the foreground would also be consistent
with the photometry and low A$_{v}$ $\approx$ 0
(Hughes et al. 1994). Given
the tendancy of main sequence late M stars to have strong 
H$\alpha$ lines, the low disk luminosity, and the N$_H$
value which is suggestive of a foreground object, it is
necessary to question the CTTS identity of Sz 84. Observations
of the Li I 6707 \AA\ line will settle the question of
whether this source is a foreground young star with
a debris disk such as St 34 in Taurus (Hartmann et al. 2005b).

\section{Discussion}
\subsection{Disks, Companions, or Extragalactic Confusion?} 
Two of the WTTS in our survey show excesses only at 24 microns.
Although we are inclined to attribute this excess to dust
emission in a disk due to analogy with CTTS, it is also
conceivable that excesses could be caused by an unresolved
nearby source. The MIPS 24 $\micron$ PSF has a FWHM of 5.7$\arcsec$,
larger than the IRAC PSFs by a factor of 2. Thus, it is
possible that a nearby source could be blended into the PSF
at the longer wavelength. However, in neither case is there
an obvious IRAC source near the position of the 24 $\micron$
source. An exact overlap with an extragalactic interloper
is also a possibility, although the areal density of these
sources is only 300 deg$^{-2}$ for fluxes $\geq$ 1 mJy (Papovich et al. 2004).
Given a random distribution of galaxies, the probability of
having one source with a superposition in the 6$\arcsec$ MIPS 24 
$\micron$ beam is 0.065\%.
Thus, in our list of 90 stars, there is about a 6\% probability of
a chance superposition or 0.138\%\ for two superpositions. 
It is also true that by selecting for
X-ray bright stars, we may also be selecting for X-ray bright
AGN which are also bright at 24 $\micron$ (Franchesini et al. 2005).
Most extragalactic sources have relatively flat spectra
between 8 and 24 microns (Rowan-Robinson et al. 2005), and the 
sensitivity of IRAC is higher than MIPS, so extragalactic
interlopers would likely be detected at the IRAC bands as well
as 24 $\micron$. Unfortunately, the brightness of the stellar
photosphere in the IRAC bands will likely drown out the
extragalactic component. It is at 24 $\micron$ for stars with
photospheric levels of $\approx$ 1 mJy that this problem is
most likely, as in our survey. Unfortunately, very high
resolution imaging at high contrast will be required to 
completely rule
out this conundrum for low luminosity WTTS disks.
A final possibility is the presence of a low
mass stellar companion that is unresolved even at optical
wavelengths. Due to the shape of the possible companion SED, it is
highly unlikely that such a source would be undetected at 8 microns.
These purported companions would have to be 3 times as bright
as the primary at 24 \micron\, but less than 10\%
as bright at 8 \micron\ to fulfill the SED requirements.
The only plausible stellar companion of this sort
would be obscured by an edge-on disk and visible only
in scattered light shortward of 10 microns, as in the case
of HV Tau C, where an edge-on disk is an optical companion
of a WTTS binary (Hartmann et al. 2005a). 
\subsection{Comparison to Disk Models}
In order to investigate the physical parameters 
of the WTTS disks identified in Section 3, we
compare their SEDs to disk models. We use the passive disk model presented by Dullemond  et al. (2001, D01 
hereafter), which is based on the Chiang $\&$ Goldreich (1999) flared disk model, but includes an evacuated inner 
cavity and an inner rim. The specific application of the Dullemond model to T Tauri star
(rather than Herbig Ae/Be stars) has been called into question recently by Vinkovic et al. (2006)
and Isella \& Natta (2005). However, in the current case, this model is only used
to provide preliminary insight into possible disk structures. It is entirely likely
that more conventional T Tauri disk models (e.g., D'Alessio et al. 2005) would also fit the data, potentially
requiring smaller inner disk holes. 
In the context of this model, the SED of an object can be constructed from the 
superposition of the flux  contributions of 4 different components: the stellar photosphere, an inner rim, the 
disk's surface layer, and the disk's  mid-plane. Figures 6-8 show the SEDs and disk models for the PMS stars Sz 76, Sz 
82, and Sz 84. As described in section 3.2, Sz 76 and Sz 82 are two borderline WTTS/CTTS, while Sz 84 is
a purported CTTS with no measurable excess at IRAC wavelengths (although it shows large 24 and 70 $\mu$m excesses). 
The stellar and disk model parameters are listed in Table 3. The stellar $T_{eff}$ and luminosity
were derived as  described in section 4.4. The stellar masses come from the evolutionary tracks by Siess et. al. (2001).
The free parameters  fitted to the disk models are: disk mass, inner and outer radii, inclination, and the power
index of the surface density of the disk (although the latter two parameters are not well constrained by the data). 
We find that for these particular models, the excesses at
IRAC wavelengths are completely dominated 
by the blackbody emission from the inner rim. Therefore, the inner radius and the inclination of the disk are 
relatively well constrained by the relative intensities and magnitude of the IRAC excesses. The total mass of the disk, 
the outer radius, and the power index are degenerate parameters only partially constrained by the 24 and 70 $\mu$m fluxes. 
These are very preliminary results, but they show that the SEDs of WTTS can be well reproduced by simple 
passive flared disk models with large inner holes. 

\subsection{Disks Are Rare Among WTTS}
If the low frequency of disks among WTTS is confirmed by the
remainder of the c2d WTTS survey, it appears that most
weak-line T Tauri stars are indeed ``naked" T Tauri stars
at circumstellar radii less than those probed by 24 $\micron$
emission.
A substantial population of stars apparently coeval with
the CTTS have fully disposed of their inner and outer
disks within a time frame of a few million years. These
results are similar to those found by Hartmann et al. (2005a)
for WTTS in Taurus-Auriga. These authors found that the
[3.6] - [4.5] versus [5.8] - [8.0] diagram clearly
separates the bulk of CTTS (IRAC excess sources) from WTTS 
(non-excess sources) with a few ambiguous cases such as
CoKu Tau/4 which has excess only beyond 15 \micron.
However, the present study
extends these conclusions beyond 8 microns to include
the outer disks probed by 24 and 70 \micron.
Another IRAC excess survey among young stars 3 - 30 Myr
in age is reported by the FEPS group (Silverstone et al. 2006). These
authors found that 5/74 sources possess infrared
excesses and four of these have ages estimated in
the 3 - 10 Myr range. Interestingly, three of their
five excess stars would be classified as CTTS based
on their H$\alpha$ EW. A few {\it Spitzer} excess
surveys of slightly older stars have also been
published recently.
Among the 23 G-type - M-type members of the TW Hya association
with an age of about 10 Myr, the 24 \micron\ excess
fraction is about 20\% (Low et al. 2005).
However, only one of the solar-type
excess objects qualifies as a CTTS based on spectroscopic criteria (Webb
et al. 1999). An even higher excess fraction (40\%)
was found in the 8 - 10 Myr Eta Cha association, but 5/6 excess
stars are CTTS (Megeath et al. 2005). A similar fraction (35\%) has
been found for the F- and G- type stars in Upper Sco-Cen (Chen et al. 2005).
Notably, our excess fraction of 6\% is considerably lower than the 20\% excess
fraction among A stars with ages of 5 Myr or less (Rieke
et al. 2005), but these sources are not sorted by 
emission line brightness as in our survey. $IRAS$ and $ISO$ studies of
WTTS mid-infrared excess are not
sensitive enough to reach the photosphere at 24 \micron\, so
they are unable to identify more than just the optically
thick disk population (e.g., Gras-Vel\'aquez \& Ray 2005).
For a similar reason, it is important to note that some faint WTTS 
may have outer disks which remain undetected by {\it Spitzer}
due to the lower than expected sensitivity of the MIPS 70 \micron\
channel and the brightness of background nebulosity at that
wavelength. Unfortunately, the time required for {\it Spitzer} to detect
the photosphere of every WTTS at 70 \micron\ would be
prohibitive. The old but nearby solar neighborhood population has been
surveyed by {\it Spitzer} at 70 $\micron$ to much
greater depth than our survey, possibly accounting
for its higher detection rate of cold disks (Bryden
et al. 2006). Based on the 24 \micron\ and IRAC excess statistics,
the frequency of excess among WTTS is lower than any
other group of young sources up to ages of 10 Myr, but
it is higher than the approximately 1\% 24 $\micron$ excess frequency
among main sequence populations (Bryden et al. 2006). We
confirm and extend the findings of previous reseachers 
that lack of significant H$\alpha$ emission among
young solar-type stars indicates a lack of disk material
even at a very tenuous level for r $<$ 3 AU. 

\subsection{Optically Thin T Tauri Star Disks?}
\par
Two of the objects in our survey have derived fractional disk
luminosities which are among the lowest yet detected
by $Spitzer$ for very young stars.
In order to have a fractional luminosity
of 1\%, an optically thick disk must be extremely flat and thin to intecept such a
small percentage of the stellar luminosity (Chen et al. 2005, Kenyon \& Hartmann 1987).
On the other hand, these disks could be optically thin, in which case
their L$_d$/L$_*$ of 0.01 
would have approximately ten times the fractional disk luminosity of 
the $\beta$ Pic disk. If the excesses of
RXJ0432.8+1735 and RXJ1603.2-3239 are not due to a chance
extragalactic superposition, they are possibly the first optically
thin disks discovered around very young T Tauri stars. At a brightness ten times that of
$\beta$ Pic, such disks should be easily visible with
high contrast observations using {\it HST}.

\subsection{Foreground Contamination by Young Main Sequence Stars}
One caveat to the above conclusion
is continuing uncertainty as to the contamination
of the WTTS sample by $\sim$ 100 Myr old field stars
(Covino et al. 1997). A recent astrometric evaluation of
ROSAT selected young stars around Tau-Aur concludes that
the sample is heterogeneous, consisting of a few actual
Tau-Aur association members mixed with Pleiades supercluster
stars, Hyades cluster member, and other foreground young stars
(Li 2005). Certainly, the low frequency of disk excess
found among our sample is more consistent with an older
population than with million year old stars (Silverstone et al. 2006;
Rieke et al. 2005; Gorlova et al. 2004).

In our survey, the WTTS with excesses are most often found within dark
clouds or adjacent to known CTTS. V836 Tauri is on the border of Lynds
1544, Sz 82 is within 10\arcmin\ of the famous emission star
RU Lup, and Sz 76 lies only 20\arcmin\ from the CTTS GQ Lup.
Even the low-excess star RXJ0432.8+1735 is located within L1551 near GG Tauri. 
Only RXJ1603.2-3239 is a relatively isolated source. Thus,
it appears that only the WTTS that are projected onto dark clouds 
and CTTS have a measureable excess. No excesses are measured for WTTS more
than one degree off-cloud.
An additional source of contaminating young stars is present for the Lupus
star formation region. The Sco-Cen region, with its numerous members
5 - 20 Myr in age, is adjacent to Lupus and potentially overlaps the
regions from which our Lupus WTTS were selected.

In order to estimate the age spread of our sample, we place our objects in the H-R diagram.
We estimate the effective temperatures directly from the spectral type of the objects using the scale 
provided by Kenyon \& Hartmann (1995). We derive the stellar luminosities by applying a bolometric
correction (appropriate for each spectral type) to the I-band magnitudes corrected for extinction 
and assuming the nominal cloud distances presumed by c2d (Taurus 140 pc, Chamaeleon 178 pc,
Lupus 150 pc, Oph 125 pc). An important {\it caveat} is that unresolved binary companions
in our sample may produce up to a factor of two overestimates of stellar luminosity.
The bolometric corrections were taken from 
Hartigan, Strom \& Strom (1994) and $A_I$ was calculated using $A_{I}= 2.76E(R-I)$. The (R-I) colors 
come from Kenyon \& Hartmann (1995). In order to evaluate the degree of model dependence of the stellar 
ages we derive, we use two different evolutionary tracks to estimate stellar ages: those provided by 
D'Antona  $\&$ Mazzitelli (1997, D97 hereafter) and those presented by Siess et al. (2001, S01 hereafter). 
We choose these two particular evolutionary tracks because they provide the appropriate mass and age range and 
are both widely used. Figure 9 shows the ages derived for our sample using both sets of evolutionary tracks
and reveals a considerable age spread. The error bars for every object have been calculated by propagating 
into the evolutionary tracks the observational uncertainties. We adopted a $T_{eff}$ uncertainty equal 
to one spectral type subclass and a luminosity uncertainty of 30$\%$ (dominated by the distance uncertainty
for individual stars).
Ages of PMS stars are very difficult to estimate due to the large observational and model uncertainties involved. 
However, we find that, (1) even though D97 and S01 evolutionary tracks show some systematic differences 
(i.e. D97 tracks yield significantly younger ages than S01 models), the relative ages agree fairly well, 
and (2) the total age spread in the sample is significantly larger than the typical (observational) 
error bar. We also find a possible decrease in the excess frequency of WTTS with
increasing derived age, although at a low significance level. According to the D97 models, 
11\% (4/37) $\pm$ 5\% of
the WTTS younger than 1 Myr have a disk, while only 2\% 
(1/46) $\pm$ 2\% of the WTTS older than 1 Myr have a disk. Similarly,
the S01 models indicate that  15\% (4/26) $\pm$ 7\% of the WTTS
younger than 3 Myrs have a disk, while only 2\% (1/57) $\pm$ 2\%
of the older stars do.  Even presuming that our stars are
at the distance of the star-forming regions with which they are associated, about a third of the stars 
have properties that are consistent with an older population.  The few sources with excess are among
the presumably younger part of the sample, although many
apparently diskless stars are also among the youngest.
In the future, careful distance determinations will be required to eliminate main sequence 
interlopers from the true WTTS sample.
\section{Conclusions}
We have detected 24 \micron\ excess emission from only 5 of 83 weak-line
T Tauri stars observed by the {\it Spitzer} ``Cores to Disks"
legacy project. The frequency for excess at $\lambda$ $<$ 10 \micron\ is
even lower (3/83); thus, two WTTS only have
excess at 24 microns. These may possess optically thin
disks. The other three have SEDs indistinguishible
from CTTS, and H$\alpha$ equivalent widths near the arbitrary 10 \AA\
dividing line between CTTS and WTTS. According to our results and
those of Hartmann et al. (2005a), solar-type
pre-main sequence stars seem to predominantly be either diskless
or surrounded by optically thick disks. There is no large population
of optically thin disks around WTTS.
Based on these preliminary
results, we conclude that both optically thick and optically
thin warm disks are rare among young solar-type stars with
low H$\alpha$ emission. This finding suggests that some stars
may pass through the Class II stage in less than 1 - 3 Myr. {\it Spitzer} 
observations at greater depth are required to determine whether the incidence of
cold outer disks (excess at $\geq$ 70 $\micron$) for WTTS 
is consistent with the field. Further work is also required to
determine the fraction of contaminating ZAMS foreground stars in
current WTTS lists. A final {\it caveat} is that frequencies of
apparently optically thin 24 $\micron$ excess among the WTTS population  
are low enough so that confusion with background galaxies in
the 6$\arcsec$ {\it Spitzer} beam cannot be ruled out as the source of the
excess for individual objects. 

\acknowledgements{We acknowledge the use of Dr. C. Dullemond's cgplus
publically available disk models in Dullemond et al. (2001).
We thank the Lorentz Center of the Leiden
Observatory for hosting a meeting in summer 2005 at which
much of this work was performed.
Support for this work, which is part of the {\it Spitzer} Legacy
Science Program, was provided by NASA through contracts 1224608,
1230782, and 1230799 issued by the Jet Propulsion Laboratory,
California Institute of Technology under NASA contract 1407. This
publication makes use of data products from the Two Micron All Sky
Survey, which is a joint project of the University of Massachusetts
and the Infrared Processing and Analysis Center funded by
NASA and the National Science Foundataion. We also acknowledge
use of the SIMBAD database.}

\vfil\eject

\clearpage
\thispagestyle{empty}
\begin{deluxetable}{lrrccl}
\tabletypesize{\tiny}
\rotate
\tiny
\tablecolumns{11}
\tablecaption{{\it Spitzer} Observations}
\tablehead{
\colhead{ID} & \colhead{RA} & \colhead{Dec} & IRAC Obs Date & MIPS Obs Date & Notes\tablenotemark{1}
}
\startdata
NTTS032641+2420 & 52.40987 & 24.51053 & 2004-08-16  & 2004-09-19  &    \\
RXJ0405.3+2009 & 61.33163 & 20.15711  & 2004-09-07  & 2004-09-22  &    \\
NTTS040234+2143 & 61.37862 & 21.85294 & 2005-02-22  & 2004-09-25  &    \\
RXJ0409.2+1716 & 62.32083 & 17.26892  & 2004-09-07  & 2004-09-25  &    \\
RXJ0409.8+2446 & 62.46304 & 24.77253  & 2004-09-08  & 2004-09-25  &    \\
RXJ0412.8+1937 & 63.21100 & 19.61614  & 2004-09-07  & 2004-09-25  & IRAC C2 with RXJ0405.3+2009 \\
NTTS041559+1716 & 64.71542 & 17.38792 & 2004-09-07  & 2004-09-25  &    \\
RXJ0420.3+3123 & 65.10050 & 31.38992  & 2004-09-07  & 2004-09-19  &    \\
RXJ0424.8+2643a & 66.20433 & 26.71956 & 2004-09-07  & 2004-09-25  &    \\
RXJ0432.8+1735 & 68.22179 & 17.59269  & 2004-09-07  & 2004-09-25  &    \\
RXJ0437.4+1851a & 69.36192 & 18.85744 & 2004-09-07  & 2005-02-25  &    \\
RXJ0438.2+2023 & 69.55433 & 20.37975  & 2004-09-07  & 2005-02-25  & IRAC C2 with RXJ0437.4+1851a \\
RXJ0438.6+1546 & 69.66279 & 15.77047  & 2004-09-07  & 2005-03-03  &    \\
RXJ0439.4+3332a & 69.85604 & 33.54572 & 2004-09-07  & 2005-02-26  &    \\
RXJ0452.5+1730 & 73.12812 & 17.50714  & 2004-09-07  & 2005-02-26  &    \\
RXJ0452.8+1621 & 73.20896 & 16.36922  & 2004-09-07  & 2005-02-26  & IRAC C2, MIPS C2 with RXJ0452.5+1730 \\
RXJ0457.2+1524 & 74.32358 & 15.41928  & 2004-09-07  & 2004-10-13  &    \\
RXJ0457.5+2014 & 74.37775 & 20.24158  & 2004-09-07  & 2004-10-13  &    \\
RXJ0458.7+2046 & 74.66558 & 20.77892  & 2004-09-07  & 2004-10-13  & IRAC C2 with RXJ0457.5+2014  \\    
RXJ0459.7+1430 & 74.94237 & 14.51539  & 2004-09-07  & 2004-10-13  & IRAC C2 with RXJ0457.2+1524  \\
V836Tau        & 75.77746 & 25.38881  & 2004-09-07  & 2004-09-25  &    \\
RXJ0902.9-7759 & 135.71379 & -77.99297 & 2004-07-04  & 2005-03-06  &    \\
RXJ0935.0-7804 & 143.73350 & -78.07203 & 2004-06-10  & 2004-07-09  &    \\
RXJ0942.7-7726 & 145.70675 & -77.44464 & 2004-06-10  & 2004-07-09  & IRAC C2 with RXJ0935.0-7804 \\
RXJ1001.1-7913 & 150.28637 & -79.21872 & 2004-05-01  & 2004-07-09  &    \\
RXJ1150.4-7704 & 177.61787 & -77.07722 & 2004-05-26  & 2005-02-27  &    \\
RXJ1216.8-7753 & 184.19137 & -77.89258 & 2004-05-26  & 2005-03-06  & IRAC C3 with RXJ1150.4-7704; MIPS C2 with RXJ 1204.6-7731  \\
RXJ1219.7-7403 & 184.93204 & -74.06589 & 2004-07-21  & 2005-03-06  &    \\
RXJ1220.4-7407 & 185.09071 & -74.12758 & 2004-07-21  & 2005-03-06  & IRAC C2 with RXJ1219.7-7403 \\
RXJ1239.4-7502 & 189.83850 & -75.04419 & 2004-07-21  & 2005-03-06  & IRAC C3 with RXJ1219.7-7403, MIPS C2 with RXJ1220.4-7407   \\
RXJ1507.6-4603 & 226.90725 & -46.05433 & 2004-07-20  & 2005-03-09  &    \\
RXJ1508.6-4423 & 227.15721 & -44.38806 & 2004-07-23  & 2005-03-09  & MIPS C2 with RXJ1507.6-4603  \\
RXJ1511.6-3550 & 227.90400 & -35.84492 & 2004-08-13  & 2005-03-05  &    \\
RXJ1515.8-3331 & 228.93904 & -33.53325 & 2004-07-20  & 2005-03-05  &    \\
RXJ1515.9-4418 & 228.96975 & -44.30481 & 2004-07-23  & 2005-03-09  & IRAC C2 with RXJ1508.6-4423 \\
RXJ1516.6-4406 & 229.15263 & -44.12233 & 2004-07-23  & 2005-03-09  & IRAC C3 with RXJ1508.6-4423 \\
RXJ1518.9-4050 & 229.72008 & -40.84800 & 2004-07-24  & 2005-03-06  &    \\
RXJ1519.3-4056 & 229.81667 & -40.93542 & 2004-07-24  & 2005-03-06  & IRAC C2, MIPS C2 with RXJ1518.9-4050   \\
RXJ1522.2-3959 & 230.54842 & -39.99747 & 2004-07-24  & 2005-03-06  & IRAC C3, MIPS C3 with RXJ1518.9-4050 \\
RXJ1523.4-4055 & 230.85654 & -40.92964 & 2004-07-24  & 2005-03-06  & IRAC C4, MIPS C4 with RXJ1518.9-4050 \\
RXJ1523.5-3821 & 230.87667 & -38.35797 & 2004-07-23  & 2005-02-27  &    \\
RXJ1524.0-3209 & 231.01271 & -32.16411 & 2004-07-23  & 2004-08-06  &    \\
RXJ1524.5-3652 & 231.13483 & -36.86742 & 2004-07-24  & 2005-03-05  &    \\
RXJ1525.5-3613 & 231.38817 & -36.22964 & 2004-07-24  & 2005-03-05  & IRAC C2, MIPS C2 with RXJ1524.5-3652 \\
RXJ1526.0-4501 & 231.49850 & -45.02103 & 2004-07-24  & 2005-03-09  &    \\
RXJ1538.0-3807 & 234.51100 & -38.12306 & 2004-08-15  & 2005-03-09  &    \\
RXJ1538.6-3916 & 234.65946 & -39.28203 & 2004-08-15  & 2005-02-27  & IRAC C2 with RXJ1538.0-3807   \\
RXJ1538.7-4411 & 234.67942 & -44.19650 & 2004-07-28  & 2005-03-09  &    \\
RXJ1540.7-3756 & 235.17150 & -37.93847 & 2004-08-15  & 2005-03-09  &    \\
RXJ1543.1-3920 & 235.77600 & -39.33872 & 2004-08-15  & 2005-03-09  & IRAC C4, MIPS C2 with RXJ1538.0-3807 \\
RXJ1546.7-3618 & 236.67171 & -36.31311 & 2004-08-12  & 2005-03-05  &    \\
RXJ1547.7-4018 & 236.92400 & -40.30742 & 2004-08-18  & 2005-03-09  &    \\
Sz76 & 237.37808 & -35.83094           & 2004-08-12  & 2005-03-05  & IRAC C2 with RXJ1546.7-3618 \\
RXJ1550.0-3629 & 237.49667 & -36.49928 & 2004-08-12  & 2005-03-05  & IRAC C3 with RXJ1546.7-3618   \\
Sz77 & 237.94563 & -35.94556           & 2004-08-12  & 2005-03-05  & IRAC C4 with RXJ1546.7-3618, MIPS C2 with RXJ1550.0-3629 \\
RXJ1552.3-3819 & 238.08133 & -38.32536 & 2004-08-12  & 2005-03-06  & IRAC C2 with Sz 82  \\
RXJ1554.9-3827 & 238.72025 & -38.46572 & 2004-08-12  & 2005-03-06  & IRAC C3 with Sz 82, MIPS C2 with RXJ1552.3-3819  \\
RXJ1555.4-3338 & 238.85921 & -33.63978 & 2004-08-12  & 2005-03-05  &    \\
RXJ1555.6-3709 & 238.89075 & -37.16142 & 2004-08-12  & 2005-03-07  & IRAC C4 with Sz 82   \\
Sz81 & 238.95958 & -38.02581           & 2004-08-12  & 2005-03-09  & IRAC C5 with Sz 82   \\
RXJ1556.1-3655 & 239.00875 & -36.92450 & 2004-08-12  & 2005-03-05  & IRAC C6 with Sz 82  \\
Sz82 & 239.03838 & -37.93492           & 2004-08-12  & 2004-08-02  &    \\
Sz84 & 239.51050 & -37.60072           & 2004-08-12  & 2005-03-07  & IRAC C7 with Sz 82  \\
RXJ1559.0-3646 & 239.74917 & -36.77239 & 2004-08-12  & 2005-03-09  &    \\
sz129 & 239.81862 & -41.95283          & 2004-09-03  & 2005-03-09  &    \\
RXJ1601.2-3320 & 240.28733 & -33.33725 & 2004-08-13  & 2005-03-07  &    \\
RXJ1602.0-3613 & 240.49658 & -36.21542 & 2004-08-12  & 2005-03-07  & IRAC C3 with RXJ1559.0-3646 \\
RXJ1603.2-3239 & 240.79921 & -32.65561 & 2004-08-13  & 2005-03-07  &    \\
RXJ1603.8-3938 & 240.96875 & -39.65036 & 2004-09-03  & 2005-03-09  &    \\
RXJ1604.5-3207 & 241.12733 & -32.12464 & 2004-08-13  & 2005-03-07  & IRAC C2 with RXJ1603.2-3239  \\
RXJ1605.6-3837 & 241.38871 & -38.62919 & 2004-08-12  & 2005-03-09  &    \\
RXJ1607.2-3839 & 241.80708 & -38.65661 & 2004-08-12  & 2005-03-09  & IRAC C2 with RXJ1605.6-3837  \\
RXJ1610.1-4016 & 242.51992 & -40.27006 & 2004-08-12  & 2005-03-09  &    \\
PZ99$\_$154920-260005 & 237.33750 & -26.00172 & 2004-08-13 & 2005-03-05 &    \\
PZ99$\_$155506.2 & 238.77600 & -25.35283 & 2004-08-13  & 2005-03-05  &    \\
PZ99$\_$155702-195 & 239.25975 & -19.84497 & 2004-08-12 & 2004-08-04  &    \\
PZ99$\_$160151-244 & 240.46454 & -24.75692 & 2004-08-12 & 2005-03-05  &    \\
PZ99$\_$160158.2 & 240.49258 & -20.13669 & 2004-08-12  & 2004-08-04  &    \\
PZ99$\_$160253.9-2 & 240.72483 & -20.38000 & 2004-08-12 & 2004-08-04  &    \\
PZ99$\_$160550-253 & 241.46104 & -25.55378 & 2004-08-12 & 2005-03-05  &    \\
PZ99$\_$160843-260 & 242.18083 & -26.03800 & 2004-08-12 & 2005-03-05  &    \\
PZ99$\_$161019.1 & 242.57992 & -25.04169   & 2004-08-12   & 2005-03-05  &    \\
WaOph1 & 242.78708 & -19.07967         & 2004-08-12  & 2005-03-05  &    \\
WaOph2 & 242.99696 & -19.11478         & 2004-08-12  & 2005-03-05  & IRAC C2 with WaOph1 \\
RXJ1612.0-1906a & 242.99696 & -19.11478 & 2004-08-12 & 2005-03-05  & in FOV of WaOph2   \\
RXJ1612.1-1915 & 243.02217 & -19.25550 & 2004-08-12  & 2005-03-05  & IRAC C3 with WaOph1 \\
RXJ1613.0-4004 & 243.33112 & -40.08633 & 2004-08-12  & 2005-03-09  &    \\
RXJ1615.3-3255 & 243.83429 & -32.91808 & 2004-08-13  & 2005-03-07  &    \\
NTTS162649-2145 & 247.45288 & -21.86994 & 2004-08-13  & 2005-03-10 &    \\
WaOph6 & 252.19008 & -14.27664 & 2004-08-16  & 2005-03-10  &    \\
\enddata
\tablenotetext{1}{The nomenclature "C\#" refers to a {\it Spitzer} cluster target in which
one observing template is applied to "\#" adjacent sources}
\end{deluxetable}

\begin{deluxetable}{lrrrrrrcccccc}
\rotate
\tablecolumns{11}
\tablecaption{2MASS and {\it Spitzer} Photometry }
\tablehead{
\colhead{ID} & \colhead{V} & \colhead{R} & \colhead{I} &
\colhead{J} & \colhead{H} & \colhead{K$_s$} & 
\colhead{F$_{3.6}$} & \colhead{F$_{4.5}$} & \colhead{F$_{5.8}$}  & 
\colhead{F$_{8.0}$} & \colhead{F$_{24}$} &\colhead{F$_{70}$} \\ 
\colhead{}  & \multicolumn{6}{c}{mag} & \multicolumn{6}{c}{mJy\tablenotemark{1}}} 
\startdata
NTTS032641+2420  &  12.20  &  11.64  &  11.13  &  10.32  &  9.86  &  9.70  & 37 & 25 & 15 &  9 & 0.9 &  - \\ 
RXJ0405.3+2009  &  10.67  &  9.96  &  9.41  &  8.69  &  8.19  &  8.09  & 150 & 100 & 67 & 38 & 4.3 &  - \\ 
NTTS040234+2143  &  14.77  &  13.72  &  12.31  &  10.95  &  10.29  &  10.06  & 31 & 20 & 14 &  9 & 0.8 &  - \\ 
RXJ0409.2+1716  &  13.44  &  12.11  &  11.15  &  9.96  &  9.25  &  9.05  & 69 & 48 & 32 & 17 & 1.4 &  - \\ 
RXJ0409.8+2446  &  13.51  &  12.55  &  11.35  &  10.10  &  9.45  &  9.25  & 55 & 40 & 26 & 15 & 1.9 &  - \\ 
RXJ0412.8+1937  &  12.47  &  11.68  &  10.85  &  9.99  &  9.43  &  9.24  & 52 & 34 & 22 & 12 & 1.3 &  - \\ 
NTTS041559+1716  &  12.23  &  11.56  &  10.88  &  10.03  &  9.42  &  9.27  & 57 & 37 & 24 & 14 & 1.2 &  - \\ 
RXJ0420.3+3123  &  12.60  &  11.96  &  11.30  &  10.45  &  9.88  &  9.73  & 39 & 24 & 15 &  9 & 0.9 &  - \\ 
RXJ0424.8+2643a  &  11.35  &  10.56  &  9.73  &  8.61  &  7.99  &  7.78  & 191 & 152 & 96 & 54 & 5.3 &  - \\ 
RXJ0432.8+1735  &  13.66  &  12.60  &  11.32  &  10.00  &  9.23  &  9.02  & 84 & 47 & 37 & 20 & 14.3 &  - \\ 
RXJ0437.4+1851a  &  00.00  &  00.00  &  00.00  &  9.42  &  8.56  &  8.67  & 101 & 61 & 44 & 24 & 2.8 &  - \\ 
RXJ0438.2+2023  &  12.18  &  11.52  &  10.90  &  10.07  &  9.53  &  9.36  & 54 & 35 & 21 & 13 & 1.1 &  - \\ 
RXJ0438.6+1546  &  10.89  &  10.31  &  9.73  &  8.90  &  8.36  &  8.24  & 134 & 82 & 58 & 34 & 4.0 &  - \\ 
RXJ0439.4+3332a  &  11.54  &  10.79  &  10.13  &  9.18  &  8.57  &  8.42  & 124 & 81 & 51 & 29 & 2.9 &  - \\ 
RXJ0452.5+1730  &  11.97  &  11.08  &  10.58  &  9.97  &  9.41  &  9.25  & 56 & 34 & 23 & 14 & 1.3 &  - \\ 
RXJ0452.8+1621  &  11.74  &  10.81  &  10.05  &  9.10  &  8.48  &  8.28  & 130 & 97 & 61 & 36 & 3.7 &  - \\ 
RXJ0457.2+1524  &  10.21  &  9.67  &  9.13  &  8.38  &  7.91  &  7.75  & 250 & 135 & 95 & 54 & 6.5 &  - \\ 
RXJ0457.5+2014  &  11.34  &  10.73  &  10.15  &  9.28  &  8.82  &  8.69  & 93 & 63 & 41 & 23 & 2.1 &  - \\ 
RXJ0458.7+2046  &  11.95  &  11.05  &  10.43  &  9.59  &  8.96  &  8.80  & 81 & 53 & 34 & 20 & 2.5 &  - \\ 
RXJ0459.7+1430  &  11.71  &  11.10  &  10.53  &  9.66  &  9.09  &  8.95  & 74 & 47 & 31 & 18 & 1.5 &  - \\ 
V836Tau  &  13.99  &  12.93  &  11.74  &  9.91  &  9.08  &  8.60  & 120 & 127 & 107 & 116 & 163 &  120 \\ 
RXJ0902.9-7759  &  14.01  &  12.90  &  11.51  &  10.10  &  9.45  &  9.22  & 65 & 42 & 30 & 16 & 2.2 &  - \\ 
RXJ0935.0-7804  &  13.79  &  12.66  &  11.21  &  9.79  &  9.13  &  8.89  & 93 & 63 & 40 & 22 & 2.3 &  - \\ 
RXJ0942.7-7726  &  13.59  &  12.64  &  11.58  &  10.37  &  9.68  &  9.46  & 47 & 33 & 21 & 12 & 1.3 &  - \\ 
RXJ1001.1-7913  &  13.23  &  12.28  &  11.22  &  10.07  &  9.38  &  9.22  & 60 & 41 & 28 & 15 & 1.6 &  - \\ 
RXJ1150.4-7704  &  11.88  &  11.20  &  10.62  &  9.71  &  9.13  &  8.97  & 82 & 50 & 32 & 18 & 1.9 &  - \\ 
RXJ1216.8-7753  &  14.03  &  12.94  &  11.61  &  10.09  &  9.46  &  9.24  & 59 & 41 & 27 & 15 & 1.8 &  - \\ 
RXJ1219.7-7403  &  12.93  &  11.96  &  11.09  &  9.75  &  9.05  &  8.86  & 87 & 58 & 41 & 22 & 2.4 &  - \\ 
RXJ1220.4-7407  &  12.47  &  11.53  &  10.64  &  9.26  &  8.61  &  8.37  & 140 & 94 & 62 & 35 & 4.1 &  - \\ 
RXJ1239.4-7502  &  10.32  &  9.75  &  9.22  &  8.43  &  7.95  &  7.78  & 194 & 134 & 88 & 51 & 5.3 &  - \\ 
RXJ1507.6-4603  &  11.69  &  11.11  &  10.55  &  9.82  &  9.21  &  9.10  & 68 & 46 & 30 & 16 & 1.6 &  - \\ 
RXJ1508.6-4423  &  10.33  &  9.92  &  9.65  &  9.36  &  8.93  &  8.81  & 88 & 54 & 36 & 20 & 2.0 &  - \\ 
RXJ1511.6-3550  &  12.42  &  11.74  &  11.08  &  10.10  &  9.52  &  9.33  & 55 & 37 & 25 & 13 & 1.3 &  - \\ 
RXJ1515.8-3331  &  00.00  &  00.00  &  00.00  &  8.98  &  8.46  &  8.38  & 118 & 80 & 52 & 28 & 3.3 &  - \\ 
RXJ1515.9-4418  &  12.47  &  11.81  &  11.17  &  10.18  &  9.57  &  9.45  & 47 & 29 & 20 & 12 & 1.1 &  - \\ 
RXJ1516.6-4406  &  11.99  &  11.35  &  10.77  &  9.90  &  9.32  &  9.19  & 53 & 37 & 25 & 14 & 1.4 &  - \\ 
RXJ1518.9-4050  &  10.90  &  10.39  &  9.88  &  9.14  &  8.66  &  8.55  & 112 & 68 & 47 & 27 & 2.6 &  - \\ 
RXJ1522.2-3959  &  12.08  &  11.42  &  10.77  &  9.91  &  9.30  &  9.10  & 64 & 44 & 28 & 17 & 1.6 &  - \\ 
RXJ1523.4-4055  &  11.95  &  11.34  &  10.76  &  9.96  &  9.39  &  9.26  & 57 & 37 & 25 & 14 & 1.2 &  - \\ 
RXJ1523.5-3821  &  14.24  &  13.18  &  11.83  &  10.40  &  9.70  &  9.50  & 48 & 33 & 23 & 13 & 1.5 &  - \\ 
RXJ1524.0-3209  &  12.36  &  11.52  &  10.60  &  9.50  &  8.82  &  8.64  & 106 & 70 & 47 & 24 & 2.7 &  - \\ 
RXJ1524.5-3652  &  10.84  &  10.35  &  10.03  &  9.55  &  9.05  &  8.93  & 81 & 48 & 32 & 18 & 1.8 &  - \\ 
RXJ1526.0-4501  &  10.95  &  10.48  &  10.05  &  9.44  &  8.98  &  8.90  & 75 & 48 & 32 & 18 & 1.6 &  - \\ 
RXJ1538.0-3807  &  12.32  &  11.62  &  10.95  &  10.11  &  9.57  &  9.38  & 52 & 35 & 23 & 13 & 1.3 &  - \\ 
RXJ1538.6-3916  &  11.60  &  11.00  &  10.41  &  9.59  &  9.01  &  8.85  & 80 & 50 & 32 & 18 & 1.8 &  - \\ 
RXJ1538.7-4411  &  10.47  &  9.99  &  9.51  &  8.80  &  8.34  &  8.21  & 134 & 88 & 59 & 33 & 4.7 &  - \\ 
RXJ1540.7-3756  &  12.19  &  11.45  &  10.94  &  9.93  &  9.32  &  9.19  & 61 & 37 & 24 & 15 & 1.3 &  - \\ 
RXJ1546.7-3618  &  11.39  &  10.85  &  10.28  &  9.49  &  8.95  &  8.78  & 97 & 56 & 40 & 23 & 2.0 &  - \\ 
RXJ1547.7-4018  &  11.06  &  10.51  &  10.00  &  9.29  &  8.81  &  8.66  & 87 & 60 & 42 & 22 & 1.9 &  - \\ 
Sz76  &  15.15  &  13.98  &  12.45  &  10.96  &  10.28  &  10.02  & 34 & 26 & 22 & 28 & 64 &  120 \\ 
RXJ1550.0-3629  &  00.00  &  00.00  &  00.00  &  9.56  &  9.02  &  8.88  & 76 & 52 & 36 & 20 & 1.9 &  - \\ 
Sz77  &  12.11  &  11.29  &  10.50  &  9.44  &  8.59  &  8.27  & 223 & 181 & 160 & 189 & 343 &  110 \\ 
RXJ1552.3-3819  &  13.08  &  12.23  &  11.41  &  10.36  &  9.68  &  9.53  & 45 & 30 & 19 & 11 & 1.1 &  - \\ 
RXJ1554.9-3827  &  13.43  &  12.53  &  11.57  &  10.41  &  9.74  &  9.57  & 47 & 31 & 21 & 12 & 1.2 &  - \\ 
RXJ1555.6-3709  &  12.39  &  11.65  &  10.94  &  9.96  &  9.33  &  9.16  & 64 & 40 & 27 & 15 & 1.2 &  - \\ 
Sz81  &  14.87  &  13.53  &  11.78  &  10.18  &  9.53  &  9.17  & 89 & 67 & 72 & 71 & 79.4 &  110 \\ 
RXJ1556.1-3655  &  13.37  &  12.57  &  11.65  &  10.40  &  9.60  &  9.30  & 82 & 72 & 55 & 62 & 171 &  154 \\ 
Sz82  &  11.85  &  11.01  &  10.13  &  8.78  &  8.09  &  7.74  & 315 & 284 & 277 & 380 & 648 &  987 \\ 
Sz84  &  16.13  &  14.70  &  12.88  &  10.93  &  10.20  &  9.85  & 42 & 29 & 20 & 12 & 20.9 &  244 \\ 
RXJ1559.0-3646  &  13.50  &  12.51  &  11.31  &  10.14  &  9.46  &  9.28  & 67 & 43 & 30 & 17 & 1.5 &  - \\ 
Sz129  &  12.93  &  12.07  &  11.13  &  9.93  &  9.08  &  8.61  & 170 & 159 & 130 & 165 & 274 &  254 \\ 
RXJ1601.2-3320  &  11.06  &  10.50  &  9.91  &  9.03  &  8.55  &  8.53  & 108 & 68 & 46 & 30 & 3.2 &  - \\ 
RXJ1602.0-3613  &  11.87  &  11.22  &  10.53  &  9.60  &  8.97  &  8.85  & 85 & 61 & 38 & 22 & 2.1 &  - \\ 
RXJ1603.2-3239  &  12.71  &  11.90  &  11.05  &  9.98  &  9.29  &  9.12  & 67 & 46 & 32 & 18 & 8.3 &  - \\ 
RXJ1607.2-3839  &  12.72  &  11.85  &  10.93  &  9.69  &  8.96  &  8.88  & 88 & 55 & 39 & 23 & 2.4 &  - \\ 
RXJ1610.1-4016  &  11.18  &  10.61  &  10.06  &  9.34  &  8.80  &  8.62  & 98 & 69 & 46 & 26 & 2.8 &  - \\ 
PZ99$\_$154920-260  &  00.00  &  10.35  &  9.71  &  8.64  &  8.13  &  7.91  & 184 & 117 & 80 & 47 & 4.4 &  - \\ 
PZ99$\_$155506.2  &  12.30  &  11.42  &  10.52  &  9.41  &  8.62  &  8.51  & 123 & 80 & 52 & 29 & 2.9 &  - \\ 
PZ99$\_$155702-195  &  00.00  &  11.01  &  10.24  &  9.18  &  8.49  &  8.37  & 141 & 91 & 61 & 34 & 3.2 &  - \\ 
PZ99$\_$160151-244  &  00.00  &  11.38  &  10.54  &  9.41  &  8.66  &  8.48  & 122 & 74 & 52 & 29 & 3.0 &  - \\ 
PZ99$\_$160158.2  &  10.45  &  9.87  &  9.27  &  8.35  &  7.81  &  7.67  & 199 & 147 & 106 & 60 & 6.2 &  - \\ 
PZ99$\_$160253.9-2  &  00.00  &  11.54  &  10.58  &  9.16  &  8.46  &  8.19  & 158 & 115 & 85 & 44 & 4.8 &  - \\ 
PZ99$\_$160550-253  &  00.00  &  10.46  &  9.99  &  9.11  &  8.59  &  8.46  & 105 & 76 & 49 & 28 & 3.1 &  - \\ 
PZ99$\_$160843-260  &  00.00  &  10.01  &  9.30  &  8.55  &  8.05  &  7.91  & 197 & 125 & 86 & 47 & 4.9 &  - \\ 
PZ99$\_$161019.1  &  11.89  &  11.06  &  10.27  &  9.25  &  8.54  &  8.36  & 121 & 83 & 57 & 32 & 3.9 &  - \\ 
WaOph1  &  11.98  &  11.11  &  10.17  &  8.76  &  7.98  &  7.69  & 282 & 174 & 117 & 66 & 7.1 &  - \\ 
WaOph2  &  11.69  &  10.96  &  10.35  &  8.98  &  8.32  &  8.09  & 127 & 102 & 69 & 38 & 5.0 &  - \\ 
RXJ1612.1-1915  &  13.57  &  12.55  &  11.47  &  10.05  &  9.08  &  8.86  & 90 & 52 & 39 & 21 & 1.8 &  - \\ 
RXJ1613.0-4004  &  00.00  &  00.00  &  00.00  &  10.47  &  9.44  &  9.11  & 72 & 43 & 31 & 18 & 1.9 &  - \\ 
RXJ1615.3-3255  &  12.04  &  11.28  &  10.54  &  9.44  &  8.78  &  8.56  & 114 & 85 & 61 & 66 & 271 &  727 \\ 
WaOph6  &  13.22  &  12.02  &  10.79  &  8.72  &  7.57  &  6.86  & 1018 & 924 & 871 & 814 & 1054 &  989 \\ 
RXJ1519.3-4056  &  11.44  &  10.84  &  10.27  &  9.55  &  9.02  &  8.83  & 90 & 58 & 37 & 21 & 2.1 &  - \\ 
RXJ1525.5-3613  &  11.64  &  11.04  &  10.44  &  9.56  &  9.00  &  8.84  & 74 & 58 & 38 & 22 & 2.1 &  - \\ 
RXJ1543.1-3920  &  12.20  &  11.48  &  10.79  &  9.85  &  9.21  &  9.10  & 62 & 43 & 29 & 16 & 1.3 &  - \\ 
RXJ1555.4-3338  &  12.48  &  11.77  &  11.06  &  10.16  &  9.54  &  9.35  & 49 & 34 & 23 & 13 & 1.6 &  - \\ 
RXJ1603.8-3938  &  11.23  &  10.58  &  9.90  &  8.94  &  8.36  &  8.22  & 149 & 93 & 63 & 35 & 3.5 &  - \\ 
RXJ1604.5-3207  &  10.81  &  10.29  &  9.79  &  9.17  &  8.69  &  8.56  & 119 & 73 & 49 & 28 & 2.7 &  - \\ 
RXJ1605.6-3837  &  14.24  &  13.22  &  12.05  &  10.79  &  10.10  &  9.90  & 37 & 23 & 16 &  9 & 1.0 &  - \\ 
RXJ1612.0-1906a  &  11.60  &  10.87  &  10.10  &  8.98  &  8.32  &  8.09  & 127 & 102 & 69 & 38 & 5.0 &  - \\ 
NTTS162649-2145  &  11.06  &  10.32  &  9.81  &  8.68  &  8.00  &  7.76  & 248 & 153 & 104 & 57 & 6.2 &  - \\ 
\enddata
\tablenotetext{1}{Photometric measurement uncertainties are estimated at 1\% for 2MASS, 15\% for IRAC and MIPS-24,
and 20\% for MIPS-70. Absolute calibration uncertainties are discussed in the text}

\end{deluxetable}

\begin{deluxetable}{lcc}
\tablewidth{0pt}
\tablecaption{Fractional Disk Luminosities}
\tablehead{\colhead{Star}&\colhead{L$_d$/L$_*$}&\colhead{CTTS/WTTS}}
\startdata
Wa Oph 6       & 0.19   & CTTS \\
RXJ1556.1-3655 & 0.10   & CTTS \\
Sz 81          & 0.05   & CTTS \\
Sz 129         & 0.11  & CTTS \\
Sz 84          & $\sim$0.0008 & CTTS? \\
RXJ1615.3-3255 & 0.08  & CTTS \\
Sz 77          & 0.11  & CTTS \\
Sz 76          & 0.10  & WTTS \\
Sz 82          & 0.12  & WTTS \\
V836 Tau       & 0.11  & WTTS \\
RXJ0432.8+1735 & $\sim$0.01  & WTTS \\
RXJ1603.2-3239 & $\sim$0.01  & WTTS \\
\enddata
\end{deluxetable}

\begin{deluxetable}{rrrrrrrrr}
\tablewidth{0pt}
\tablecaption{Model Parameters}
\tablehead{\colhead{Star}&\colhead{$L_{*}$}&\colhead{$T_{eff}$}&\colhead{$M_{*}$}&\colhead{$M_{disk}$}&\colhead{$R_{in}$}&\colhead{$R_{out}$}&\colhead{Incl}&\colhead{psig} \\
\colhead{}&\colhead{$(L_{\odot})$}&\colhead{$(K)$}&\colhead{$(M_{\odot})$}&\colhead{$(M_{\odot})$}&\colhead{$(AU)$}&\colhead{$(AU)$}&\colhead{(deg)}&\colhead{}} 

\startdata
Sz 76         & $0.25$ &  $3580 $  & $   0.35  $ & $ 1.2$x$10^{-3} $ &$  1.0 $ & $ 105   $ & $  10  $ & $ -2    $      \\
Sz 82         & $1.65$ &  $3850 $  & $   0.60  $ & $ 5.0$x$10^{-4} $ &$  0.5 $ & $ 150   $ & $  25  $ & $ -1.5  $      \\
Sz 84         & $0.35$ &  $3370 $  & $   0.30  $ & $ 1.5$x$10^{-4} $ &$  3.0 $ & $ 250   $ & $  0.5 $ & $ -2    $      \\
\enddata
\end{deluxetable}
\clearpage
\begin{figure}
\plotone{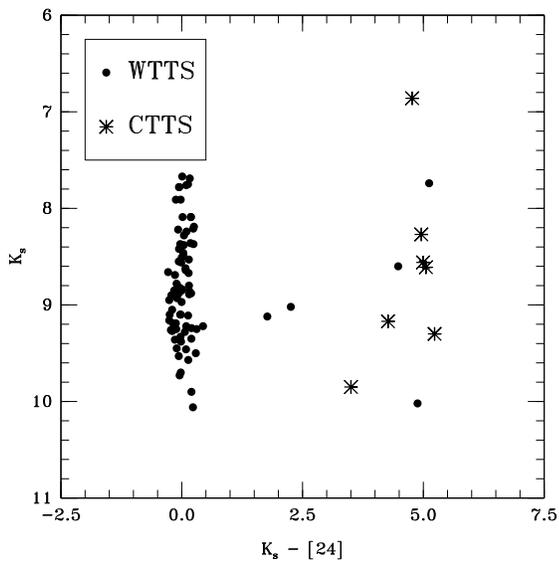}
\caption{K$_s$ versus K$_s$ - [24] for 90 young stars
with low and moderate H$\alpha$ emission}
\end{figure}

\begin{figure}
\plotone{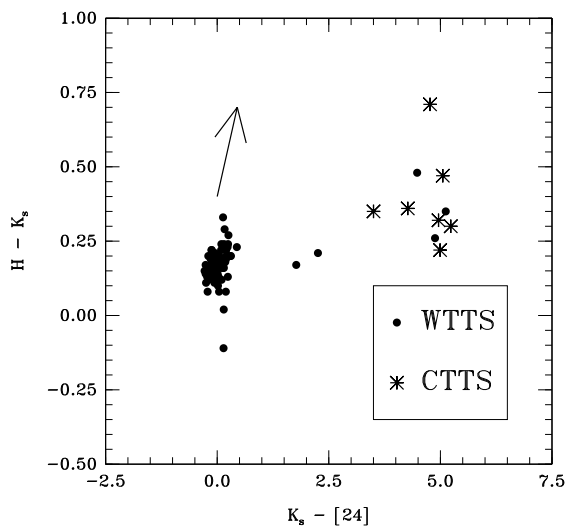}
\caption{K$_s$ - H versus K$_s$ - [24] for 90 young stars
with low and moderate H$\alpha$ emission. The reddening vector
is for A$_v$ = 5. This plot shows that high extinction sources
will not be confused with 24 $\micron$ excess sources.}
\end{figure}

\begin{figure}
\plotone{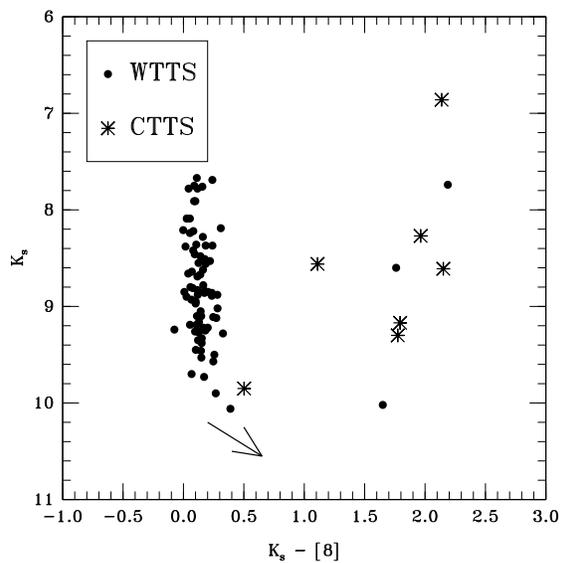}
\caption{K$_s$ versus K$_s$ - [8] for 90 young stars
with low and moderate H$\alpha$ emission. The reddening vector
is for A$_v$ = 5.}
\end{figure}

\begin{figure}
\plotone{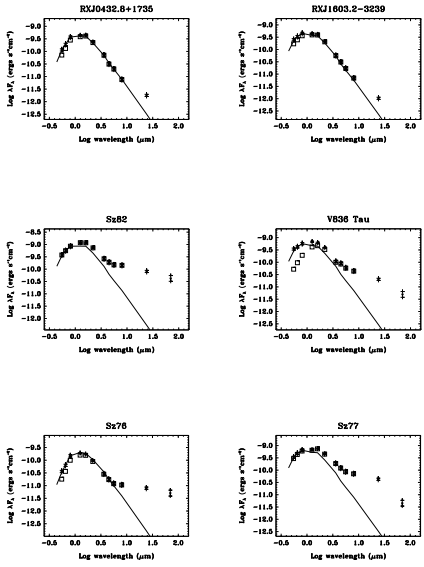}
\caption{Spectral energy distributions and model photospheres
for WTTS with excess.  Points are plotted for optical photometry
, 2MASS, IRAC, and MIPS. Dereddened fluxes are indicated by
squares.  Model photospheres based on the published spectral
types for each star are shown for comparison.}
\end{figure}

\begin{figure}
\plotone{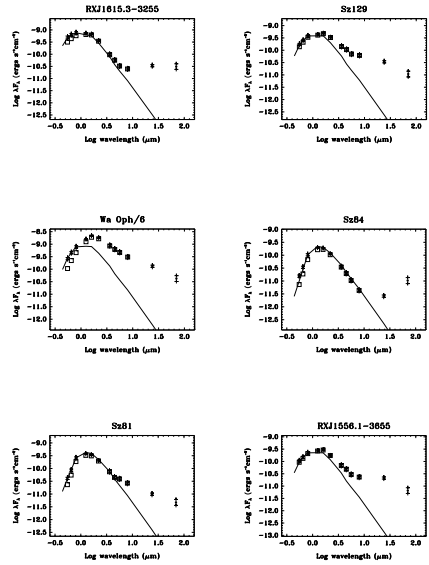}
\caption{Spectral energy distributions and model photospheres
for CTTS.  Points are plotted for optical, 2MASS,
IRAC, and MIPS photometry; dereddened fluxes are
indicated by squares. Model photospheres based on the published spectral
types for each star are shown for comparison.}
\end{figure}

\begin{figure}
\plotone{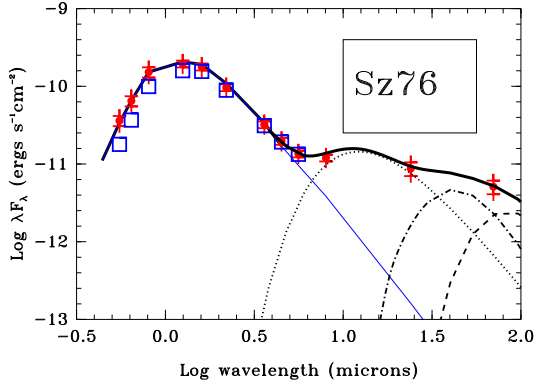}
\caption{Sz 76 SED and disk model. The bold solid line indicates the
integrated model flux. The fainter solid line shows the stellar
photosphere. The dotted curve represents the contribution from the
disk inner rim, the dash-dot curve represents the disk photosphere,
and the dashed curve represents the disk interior flux.}
\end{figure}

\begin{figure}
\plotone{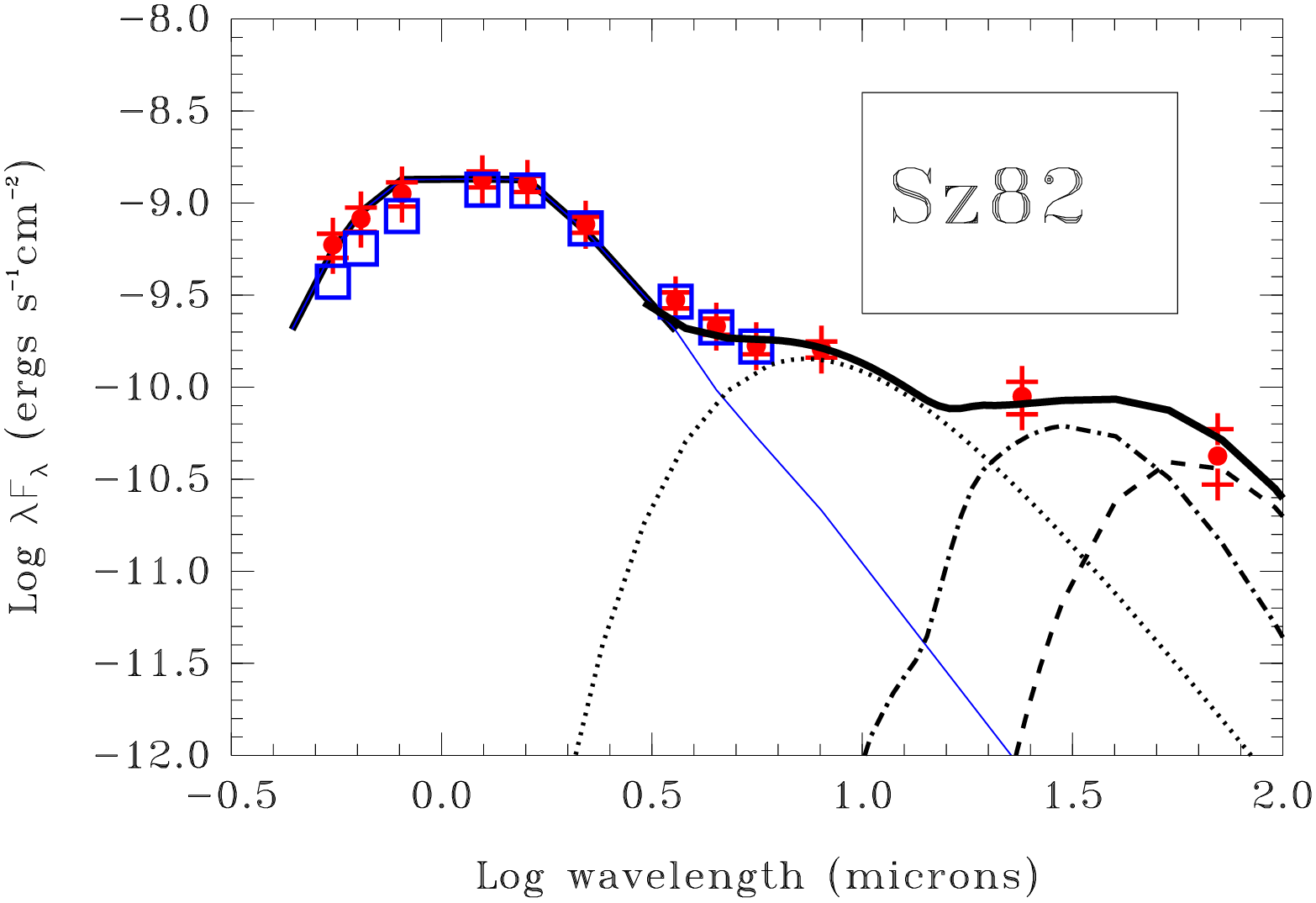}
\caption{Sz 82 SED and model as in Figure 6}
\end{figure}

\begin{figure}
\plotone{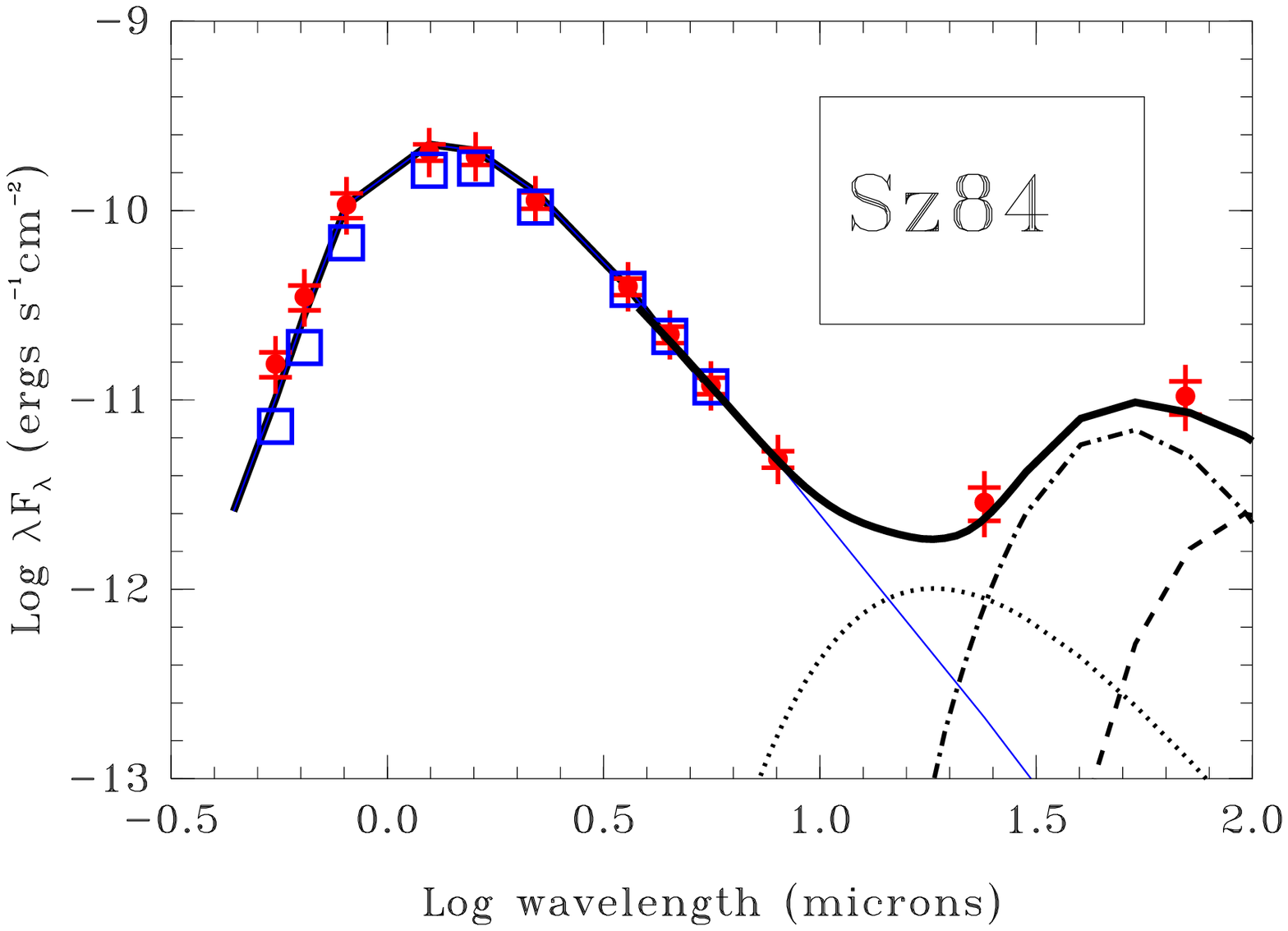}
\caption{Sz 84 SED and model as in Figure 6}
\end{figure}

\begin{figure}
\plotone{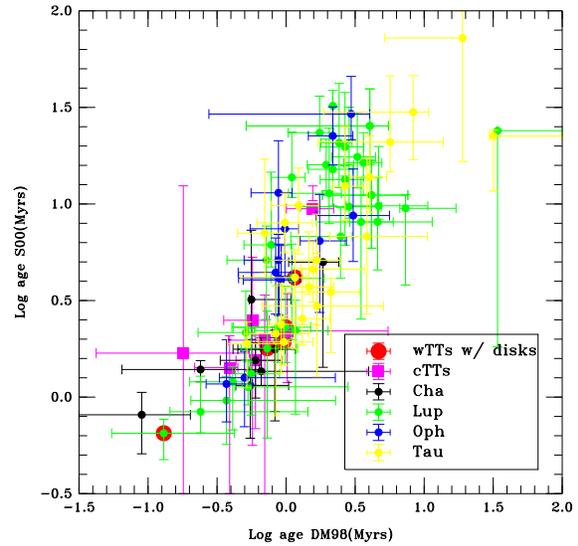}
\caption{Ages of WTTS with and without disks as derived from
the model tracks of Siess et al (2001; S01) and
D'Antona \& Mazzatelli (1997; DM97). Note that the distance
uncertainties to the stars dominate this plot.}
\end{figure}


\end{document}